\documentclass[preprint,12pt]{elsarticle}
\usepackage{float}
\usepackage{rotfloat}
\usepackage{fancyhdr}
\usepackage{url}
\usepackage{caption}
\usepackage{subcaption}
\usepackage{textcomp}
\usepackage{amsmath}
\usepackage{mathtools}
\usepackage{rotating}
\usepackage{lineno}
\usepackage{comment}
\usepackage{amssymb}
\usepackage[utf8]{inputenc}

\begin{document}
\begin{frontmatter}

\title{{\small Published in Astroparticle Physics as DOI:10.1016/j.astropartphys.2017.09.001}\vspace{2.cm}
 {\bf Spectral Calibration of the Fluorescence Telescopes of the Pierre Auger Observatory} }

\author[78]{A.~Aab}
\author[69]{P.~Abreu}
\author[50,49]{M.~Aglietta}
\author[32]{I.~Al Samarai}
\author[18]{I.F.M.~Albuquerque}
\author[1]{I.~Allekotte}
\author[8,11]{A.~Almela}
\author[65]{J.~Alvarez Castillo}
\author[77]{J.~Alvarez-Mu\~niz}
\author[41,43]{G.A.~Anastasi}
% ...

\address[1]{Centro At\'omico Bariloche and Instituto Balseiro (CNEA-UNCuyo-CONICET), San Carlos de Bariloche, Argentina}
% ...
\address[8]{Instituto de Tecnolog\'\i{}as en Detecci\'on y Astropart\'\i{}culas (CNEA, CONICET, UNSAM), Buenos Aires, Argentina}
% ...
\address[11]{Universidad Tecnol\'ogica Nacional -- Facultad Regional Buenos Aires, Buenos Aires, Argentina}
% ...
\address[18]{Universidade de S\~ao Paulo, Instituto de F\'\i{}sica, S\~ao Paulo, SP, Brazil}
% ...
\address[32]{Laboratoire de Physique Nucl\'eaire et de Hautes Energies (LPNHE), Universit\'es Paris 6 et Paris 7, CNRS-IN2P3, Paris, France}
% ...
\address[41]{Gran Sasso Science Institute (INFN), L'Aquila, Italy}
% ...
\address[43]{INFN Laboratori Nazionali del Gran Sasso, Assergi (L'Aquila), Italy}
% ...
\address[49]{INFN, Sezione di Torino, Torino, Italy}
\address[50]{Osservatorio Astrofisico di Torino (INAF), Torino, Italy}
% ...
\address[65]{Universidad Nacional Aut\'onoma de M\'exico, M\'exico, D.F., M\'exico}
% ...
\address[69]{Laborat\'orio de Instrumenta\c{c}\~ao e F\'\i{}sica Experimental de Part\'\i{}culas -- LIP and Instituto Superior T\'ecnico -- IST, Universidade de Lisboa -- UL, Lisboa, Portugal}
% ...
\address[77]{Universidad de Santiago de Compostela, Santiago de Compostela, Spain}
\address[78]{IMAPP, Radboud University Nijmegen, Nijmegen, The Netherlands}
% ...
\address[c]{also at Vrije Universiteit Brussels, Brussels, Belgium}
% ...

\cortext[cor1]{Corresponding Author. E-mail address: auger\_spokespersons@fnal.gov}

\begin{abstract}
We present a novel method to measure precisely the relative spectral response of the fluorescence telescopes of the Pierre Auger Observatory. We used a portable light source based on a xenon flasher and a monochromator to measure the relative spectral efficiencies of eight telescopes in steps of 5 nm from 280 nm to 440 nm.  Each point in a scan had approximately 2 nm FWHM out of the monochromator. Different sets of telescopes in the observatory have different optical components, and the eight telescopes measured represent two each of the four combinations of components represented in the observatory. We made an end-to-end measurement of the response from different combinations of optical components, and the monochromator setup allowed for more precise and complete measurements than our previous multi-wavelength calibrations. We find an overall uncertainty in the calibration of the  spectral response of most of the telescopes  of 1.5\% for all wavelengths; the six  oldest telescopes have larger overall uncertainties of about 2.2\%. We also report changes in physics measureables due to the change in calibration, which are generally small.
\end{abstract}


\end{frontmatter}


\section{Introduction}
The Pierre Auger Observatory~\cite{pierre2015pierre} has been designed to study the origin and the nature of ultra high-energy cosmic rays, which have energies above $10^{18}$~eV. The construction of the complete observatory following the original design finished in 2008. The observatory is located in Malarg\"ue, Argentina, and consists of two complementary detector systems, which provide independent information on the cosmic ray events. Extensive Air Showers (EAS) initiated by cosmic rays in the Earth's atmosphere are measured by the Surface Detector (SD) and the Fluorescence Detector (FD). The SD is composed of 1660 water Cherenkov detectors located mostly on a triangular array of 1.5 km spacing covering an area of roughly 3000~km$^{2}$.   The SD measures the EAS secondary particles reaching ground level~\cite{The_SD}. The FD is designed to measure the nitrogen fluorescence light produced in the atmosphere by the EAS secondary particles. The FD  is composed of 27 telescopes overlooking the SD array from four sites, Los Leones (LL), Los Morados (LM), Loma Amarilla (LA), and Coihueco (CO)~\cite{The_FD}. The SD takes data continuously, but the FD operates only on clear nights, and care is taken to avoid exposure to too much moonlight.

The energy of the primary cosmic ray is a key measurable for the science of the observatory, and the FD measurement of the energy, with lower independent systematic uncertainties, is used to calibrate the SD energy scale using events observed by both detectors. The work described here explains how the FD calibration at wavelengths across the nitrogen fluorescence spectrum has recently been improved, resulting in smaller related systematic uncertainties.

\begin{figure}[h]
  \centering
  \includegraphics[width=7.cm]{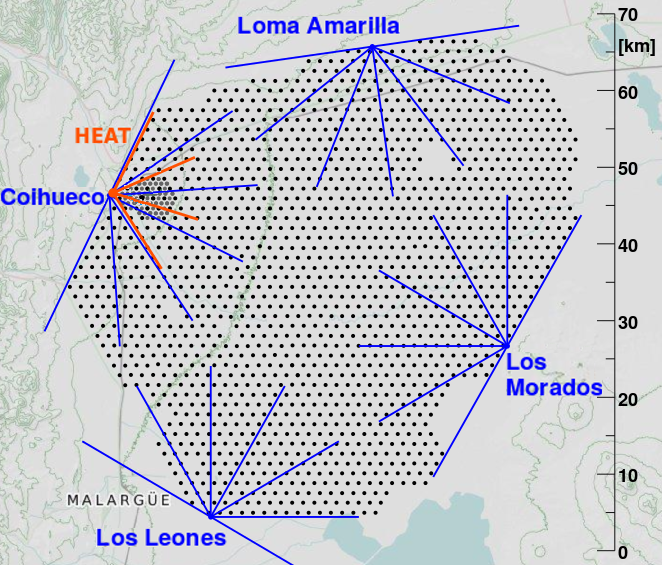}
  \caption{A schematic of the Pierre Auger Observatory where each black dot is a water Cherenkov detector. Locations of the fluorescence telescopes are shown along the perimeter of the surface detector array, where the blue lines indicate their individual field of view. The field of view of the HEAT telescopes are indicated with red lines.}
  \label{fig:PAO_layout}
\end{figure}

The buildings at the four FD sites each have six independent telescopes, and each telescope has a  30\textdegree $\times$ 30\textdegree~field of view, leading to a 180\textdegree~coverage in azimuth and from 2\textdegree~to 32\textdegree~in elevation at each building. Additionally, three specialized telescopes called HEAT~\cite{mathes2011heat} are located near Coihueco to overlook a portion of the SD array at higher elevations, from 32\textdegree~to 62\textdegree, to register EAS of lower energies. All these telescopes are housed in climate-controlled buildings, isolated from dust and day light. The layout of the observatory is shown schematically in Figure~\ref{fig:PAO_layout}.

\begin{figure}[h]
  \centering
  \includegraphics[width=7.cm]{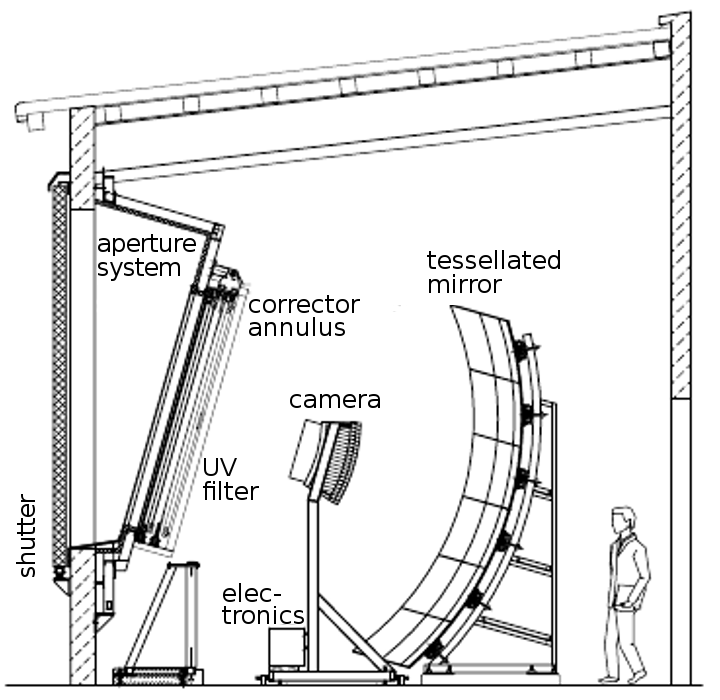}
  \caption{The optical components of an individual fluorescence telescope.}
  \label{fig:FD_Schem}
\end{figure}

Each FD telescope is composed of several optical components as shown in Figure~\ref{fig:FD_Schem}: a 2.2 m aperture diaphragm, a UV filter to reduce the background light, a Schmidt corrector annulus, a 3.5~m $\times$ 3.5~m tessellated spherical mirror, and a camera formed by an array of 440 hexagonal photomultipliers (PMT) each with a field of view of 1.5\textdegree~full angle.  Each PMT has a light concentrator approximating a  hexagonal Winston cone to reduce dead spaces between PMTs \cite{The_FD}. 

\begin{figure}[h]
  \centering
  \includegraphics[width=7.cm]{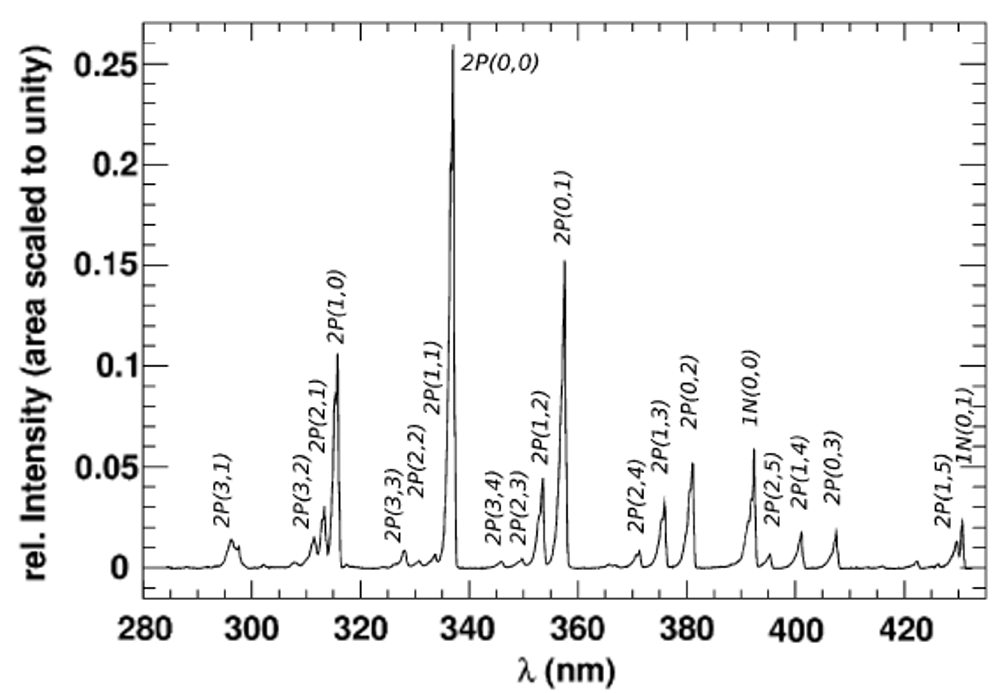}
  \caption{The nitrogen fluorescence spectrum as measured by the AIRFLY collaboration \cite{ave2008spectrally} showing the 21 major transitions.}
  \label{fig:N2_Spec}
\end{figure}

The energy calibration of the data \cite{E_Calib_2007_ICRC,Flux_Sup_2008} for the Pierre Auger Observatory, including events observed by the SD only, relies on the calibration of the FD. Events observed by both FD and SD provide the link from the FD, which is absolutely calibrated,
to the SD data. To calibrate the FD three different procedures are performed: the absolute \cite{Abs_Cal_Tech}, the relative \cite{knapik2007absolute}, and the 
spectral (or multi-wavelength) calibrations \cite{Rovero_MultiWave}. We focus here on the spectral calibration, which is a relative measurement that relates the absolute calibration performed at 365~nm to wavelengths across the nitrogen fluorescence spectrum, which is shown in Figure~\ref{fig:N2_Spec}.

To perform this measurement the drum-shaped portable light source used for the absolute calibration \cite{Abs_Cal_Tech} was adapted to emit UV light across the  wavelength range of interest. The drum light source is designed to uniformly illuminate all 440 PMTs in a single camera simultaneously when it is placed at the aperture of the FD telescope,  enabling the end-to-end calibration.

The FD response as a function of wavelength was initially calculated as a convolution of separate reflection or transmission measurements of each optical component used in the first Los Leones telescopes  \cite{matthews2003optical}.   The first end-to-end spectral calibration of the FD was performed using the drum light source with a xenon flasher and filter wheel to provide five points across the FD wavelength response \cite{Rovero_MultiWave}. This measurement represented an improvement for the energy estimation of all events observed by the Pierre Auger Observatory as it has been shown to increase the reconstructed energy of events by nearly 4\% for all energies \cite{All_Auger_ICRC_2013}. However, that result has two limitations: first, the differences in FD optical components were not measured since only one telescope was calibrated; and second, determining the FD spectral response curve using only five points involved a complicated fitting procedure, and was particularly difficult considering the large width of the filters, which resulted in  relatively large systematic uncertainties.

The aim of the work described in this paper was to measure the FD 
efficiency at many points across the nitrogen fluorescence spectrum with a reduced wavelength bite at each point, and to do it at enough telescopes to cover the different combinations of optical components making up all the telescopes within the Auger Observatory.
The spectral calibration described here proceeds in three steps. First, the relative drum emission spectrum is measured in the dark hall lab in Malarg\"ue with a specific calibration PMT, called the ``Lab-PMT'', observing the drum at a large distance, in a similar fashion to the absolute calibration of the drum; see \cite{Abs_Cal_Tech} and explanatory drawings therein. Knowing the intensity of the drum at each wavelength, we next measure the response of the FD telescopes to the output of the multi-wavelength drum over the course of several nights,  while recording data from a monitoring photodiode (PD) exposed to the narrow-band light at each point to ensure knowledge of the relative drum spectrum. 
Finally, the FD telescope response is normalized by the measured relative drum emission spectrum at every wavelength, and we evaluate the associated systematic uncertainties in the final calculation of the efficiency.

This following sections describe the measurements and analysis of data taken during March 2014: FD optical components in section 2; the new drum light source in section 3; measurements of the drum light source spectrum in section 4; calibrations performed at the FD telescopes in section 5;  FD efficiency as a function of wavelength in section 6; and final calibration results in section 7.  Effects on physics measurables due to changing calibrations are discussed in section 8.

\section{Optical Components of the Fluorescence Telescopes}
\label{sec:FD_Comp}
There are two types of mirrors used in the telescopes, and the glass used for the corrector rings was produced using two different glass-making procedures. The 12 mirrors at Los Leones and Los Morados are aluminum with a 2 mm AlMgSiO$_5$ layer glued on as the reflective surface, and the 12 mirrors at Coihueco and Loma Amarilla are composed of a borosilicate glass with a 90 nm Al layer and then a 110 nm SiO$_2$ layer (see~\cite{The_FD} for more details). Two different procedures were used to grow the borosilicate glass used in the corrector rings, both by Schott Glass Manufactures\footnote{Schott Glass, \url{http://www.us.schott.com/english/index.html}}. One type is called Borofloat 33\footnote{\label{Boro_Note}Borofloat, \url{http://www.us.schott.com/borofloat/english/attribute/optical}}, and the other is a crown glass labeled P-BK7\footnote{\label{P-BK7_Note}P-BK7, \url{http://www.schott.com/advanced_optics/us/abbe_datasheets/schott_datasheet_all_us.pdf}}, and the transmission of UV light differs for these two products.

Given the different wavelength dependencies of the above components, our aim was to measure the four combinations of mirrors and corrector rings present in the FDs. This meant calibrating at three of the four FD buildings. Table~\ref{tab:FD_Comp} shows the eight telescopes we calibrated at the three FD sites along with which components make up each telescope. Calibration of these eight telescopes gives a complete coverage of the different components and a duplicate measure of each combination.

\begin{table}[h]
\caption{List of the FD telescopes we calibrated and their respective optical components. Calibration at these eight FD telescopes gives a complete coverage of the different components and a duplicate measure of each combination. The last column indicates all other (unmeasured) telescopes with the same optical components.} 
\begin{center}
\begin{tabular}{|l|r|r|r|}
\hline
\textbf{FD telescope} & \textbf{Mirror Type} & \textbf{Corrector Ring} & \textbf{FDs with same}\\
 & & &  \textbf{components}\\
\hline
\hline
Coihueco 2    & Glass    & BK7	& ~\\
Coihueco 3    & Glass    & BK7 	& CO2/3\\\hline
Coihueco 4    & Glass    & Borofloat 33 & CO1,4-6, LA,\\
Coihueco 5    & Glass    & Borofloat 33 & HEAT\\\hline
Los Morados 4 & Aluminum & Borofloat 33 & ~\\
Los Morados 5 & Aluminum & Borofloat 33 & LM\\\hline
Los Leones 3  & Aluminum & BK7 & ~\\
Los Leones 4  & Aluminum & BK7 & LL1-6\\\hline
\hline
\end{tabular}
\end{center}
\label{tab:FD_Comp}
\end{table}

As seen in Table~\ref{tab:FD_Comp}, the telescopes CO~4/5 are the only ones that have same nominal components as those located at other FD buildings, which have different construction dates.  
It is usualy the case that optical components degrade their properties when exposed to light and ambient conditions (ageing), whose effect depends on exposure time. Even if FD telescopes are kept in climate-controlled buildings, an analysis of ageing follows. Regarding the spectral calibration, what has to be evaluated is the change in the spectral response of a given FD telescope, i.e. the shape of the response curve vs wavelength. This kind of differential degradation is not obviously seen at the FD telescopes. One way to evaluate whether there is any change in the spectral response is to track the absolute calibration done periodically at 375 nm ~\cite{pierre2015pierre, The_SD}. The absolute calibration is scaled at any given date by using the nightly relative calibration, which is done at 470 nm~\cite{pierre2015pierre, The_SD}. Because these two calibrations are done at different wavelengths, any change in the spectral response would translate in a drift of the absolute calibration with time. In Table \ref{tab:FDstarted} we show the variations of the ratio (R) of absolute calibrations performed in 2010 and 2013, where $R = (Abs_\mathrm{2013}-Abs_\mathrm{2010})/Abs_\mathrm{2013}$, along with the date of finished construction for telescopes at a given building. As seen in the table, the variations do not respond to any ageing pattern, e.g. for the oldest telescopes there is a positive variation for LL and a negative variation for CO. Moreover, the overall effect that telescopes could have in data analysis do not change the final reconstructed energy significantly (see Fig. 49 in~\cite{pierre2015pierre}). 
For these reasons, we consider that different time of telescope construction do not play a role in the spectral calibration described in this paper and, consequently, CO 4/5 can be taken as representative of LA and HEAT.

\begin{table}[h]
\caption{List of FD buildings and dates when construction was finished and operation started. $\Delta$t is the elapsed time until measurements done for this work (March 2014). R is the ratio of absolute calibrations performed in 2010 and 2013 (see text).} 
\begin{center}
\begin{tabular}{|l|r|r|r|}
\hline
\textbf{FD building} & \textbf{Built} & \textbf{$\Delta$t [yr]} & \textbf{R [\%]}\\
\hline
\hline
Los Leones  	& 5/2004    	& 9.8	& +~1.4\\\hline
Coihueco    	& 5/2004	& 9.8 	& --~1.6\\\hline
Los Morados 	& 3/2005 	& 9.0 	& --~0.5\\\hline
Loma Amarilla 	& 2/2007 	& 7.1 	& --~0.8\\\hline
HEAT 		& 9/2009	& 4.5	& ~ \\\hline
\hline
\end{tabular}
\end{center}
\label{tab:FDstarted}
\end{table}

\section{Monochromator Drum Setup}
\label{sec:Mono_Setup}

\begin{figure}[h!]
 \centering
  \includegraphics[width=13.5cm]{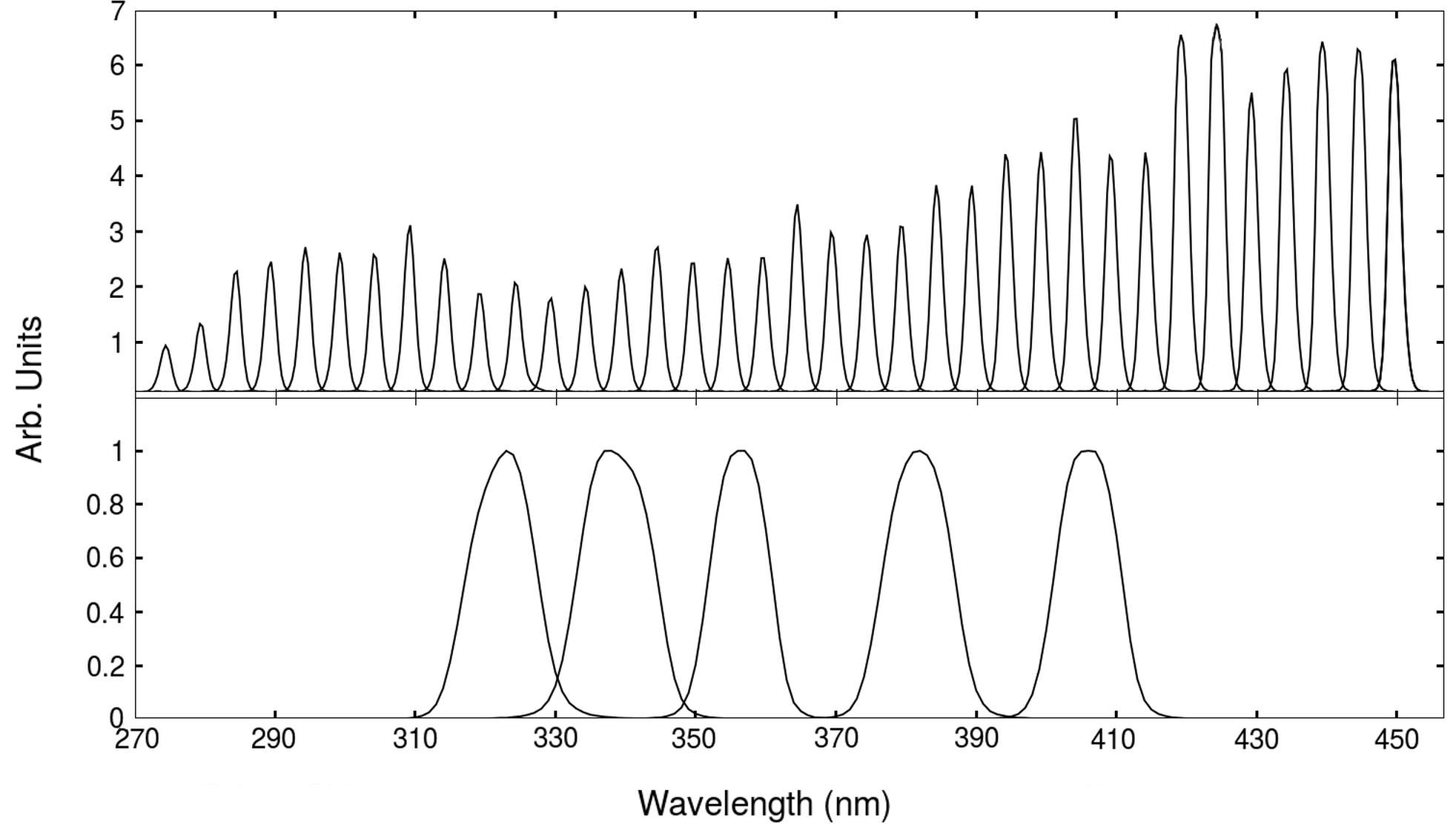}
  \caption{A comparison showing the spectral width of the output of the monochromator sampled every 5~nm (top, this work) and the notch filter spectral transmission (bottom, \cite{Rovero_MultiWave}). The y-axes are the intensity in arbitrary units for the monochromator and the normalized transmission for the notch filters.}
  \label{fig:mono_spec_notch_spec}
\end{figure}

The work described in \cite{Rovero_MultiWave} was the first in-situ end-to-end measurement of the FD efficiency as a function of wavelength. It limited the measurement to only five points across the $\sim$150~nm wide acceptance of the FDs, and the filters had a fairly wide spectral width, about $\sim$15~nm FWHM, as shown in the bottom of Figure~\ref{fig:mono_spec_notch_spec}. The large spectral width led to a complicated procedure of accounting for the width effects along the rising and falling edges of the efficiency curve  \cite{Rovero_MultiWave}. In addition, since there were only five measured points, the resulting calibration curve had to be interpolated between the points, and the original piece-wise efficiency curve \cite{matthews2003optical} was used as the starting point. In the five-point measurement \cite{Rovero_MultiWave} the efficiency was assumed to go to zero below 290~nm and above 425~nm since the filters did not extend to these wavelengths, thus the values resulting from the piece-wise convolution of the component efficiencies \cite{matthews2003optical} were the only data for wavelengths below 290~nm and above 425~nm.

These reasons are the motivation for using a monochromator to select the wavelengths out of a UV spectrum. A monochromator allows for a high resolution probe across the FD acceptance, and a far more detailed measurement can be performed. 
The top of Figure~\ref{fig:mono_spec_notch_spec} shows the output of the monochromator in 5~nm steps from 275~nm to 450~nm with a xenon flasher as the input, each step with a 2~nm FWHM. The xenon flasher is an Excelitas PAX-10 model\footnote{PAX-10 10-Watt Precision-Aligned Pulsed Xenon Light Source - \url{http://www.excelitas.com/downloads/dts_pax10.pdf}} with improved EM noise reduction and variable flash intensity. The monochromator output width was chosen to provide a reasonable compromise between wavelength resolution and the  drum intensity required for use at the FDs.

For the work described here, an enclosure housing the monochromator and xenon flasher was mounted onto the rear of the drum. The enclosure was insulated and contained a heater and associated controlling circuitry to maintain a stable 20$\pm$2~\textcelsius~temperature for monochromator reliability.

A custom 25.4~mm diameter aluminum tube was fabricated and attached to the output of the monochromator; it protrudes into the interior of the drum. At the end of the tube a 0.23~mm thick Teflon diffuser ensured that the illumination of the front face of the drum was uniform as measured with long-exposure CCD images, similar to what had been measured previously \cite{The_FD, brack2004absolute}. 

A photodiode (PD) was mounted near the output of the monochromator, but upstream of the tube that protruded into the drum, allowing for 
pulse-by-pulse monitoring of the emission spectrum from the monochromator. The monochromator and xenon flasher were controlled with the same common gateway interface (CGI) web page and calibration electronics that have been used in the absolute calibration~\cite{Abs_Cal_Tech}. Scanning  of the monochromator, triggering of the flasher, and data acquisition from monitoring devices and the FD were all fully automated using CGI code and cURL\footnote{cURL Documentation - \url{http://curl.haxx.se/}} scripts over the wireless LAN used for drum calibrations.

\section{Lab Measurements and the Drum Spectrum}
\label{sec:Lab_Meas}
To characterize the drum emission as a function of wavelength, several measurements were needed in the laboratory. 
For the one-week calibration campaign described here, four measurements were performed in the lab, two prior to any field work at the FD telescopes, one two days later and the last one at the end of the week. 

\subsection{Drum Emission}
\label{sec:Drum_Emission}
With the automated scanning of the monochromator and data acquisition we took measurements of the relative drum emission spectrum as viewed by the calibration Lab-PMT. The monitoring PD detector measured the monochromator output as described above.  The setup for these measurements had the drum at the far end of the dark hall and the Lab-PMT inside the darkbox in the calibration room, about 16~m away from the Teflon face of the drum.  See \cite{Abs_Cal_Tech} for a detailed description of the dark hall calibration setup. 

\begin{figure}[h]
 \centering
  \includegraphics[width=13.cm]{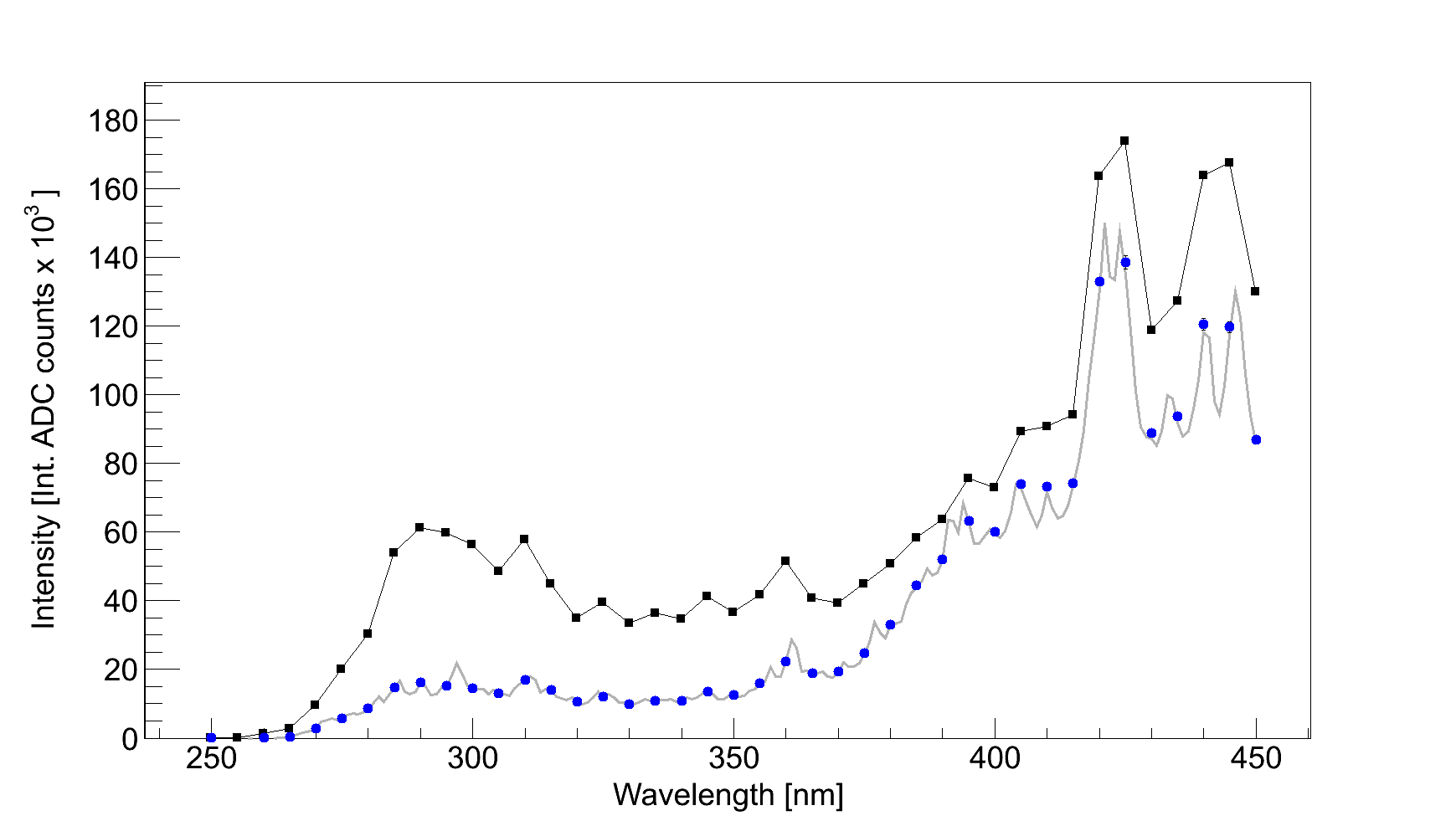}
  \caption{Drum emission spectra. Solid grey line: one of the measured spectra taken with the Lab-PMT; the line shows the average responses to 100 pulses of the drum as a function of wavelength, in steps of 1~nm. 
Blue points: the averaged drum spectrum as measured by the Lab-PMT throughout the calibration campaign; the spectrum is taken in steps of 5~nm as this is what is used to measure the FD responses; error bars are the statistical uncertainties, which are generally smaller than the plotted points. 
Black line and points: the averaged drum spectrum as measured by the monitoring photodiode (PD) throughout the calibration campaign, in steps of 5~nm.}  
\label{fig:Drum_Spec}
\end{figure}

The average response of the Lab-PMT to 100 pulses of the drum was recorded as a function of wavelength from 250~nm to 450~nm, in steps of 1~nm. The uncertainty in the average for a given wavelength was calculated as the standard deviation of the mean, $\frac{\sigma_\mathrm{Drum}}{\sqrt{100}}$. The solid grey line in Figure~\ref{fig:Drum_Spec} shows an example of one of these spectra. We took averages of the four spectra at each wavelength as the final measurement of the drum spectrum, which is shown in the same figure as blue dots. 
This final drum spectrum used measurements in steps of 5~nm corresponding to the step size used when calibrating the FD telescopes. For wavelengths between 320~nm and 390~nm, the four measurements were generally statistically consistent.  
But for wavelengths at the low and high ends of the spectrum there was disagreement; section~\ref{sec:Lab_Unc} explains how we introduce a systematic uncertainty to account for this disagreement.

For each of these four spectra measured with the Lab-PMT there are data from the monitoring PD. The monitoring PD data were handled in the same way; we made an average of the four spectra recorded by the PD and an associated error based on the spread of the four measurements.  These data are shown in Figure~\ref{fig:Drum_Spec} as  black line and points.   

\subsection{Lab-PMT Quantum Efficiency}
\label{sec:QE}
A measurement of the quantum efficiency (QE) of the Lab-PMT, which is  used to measure the relative drum emission spectrum, has to be performed to measure the relative response of a given FD telescope at different wavelengths. The method used here is similar to what was done previously \cite{Rovero_MultiWave} except, instead of a DC deuterium lamp, we used the xenon flasher as the UV light source into the monochromator. 
For the work reported here we only needed a relative measurement of the QE, and so several uncertainties associated with an absolute QE measurement are not included in this work. 

\begin{figure}[h]
 \centering
  \includegraphics[width=11.cm]{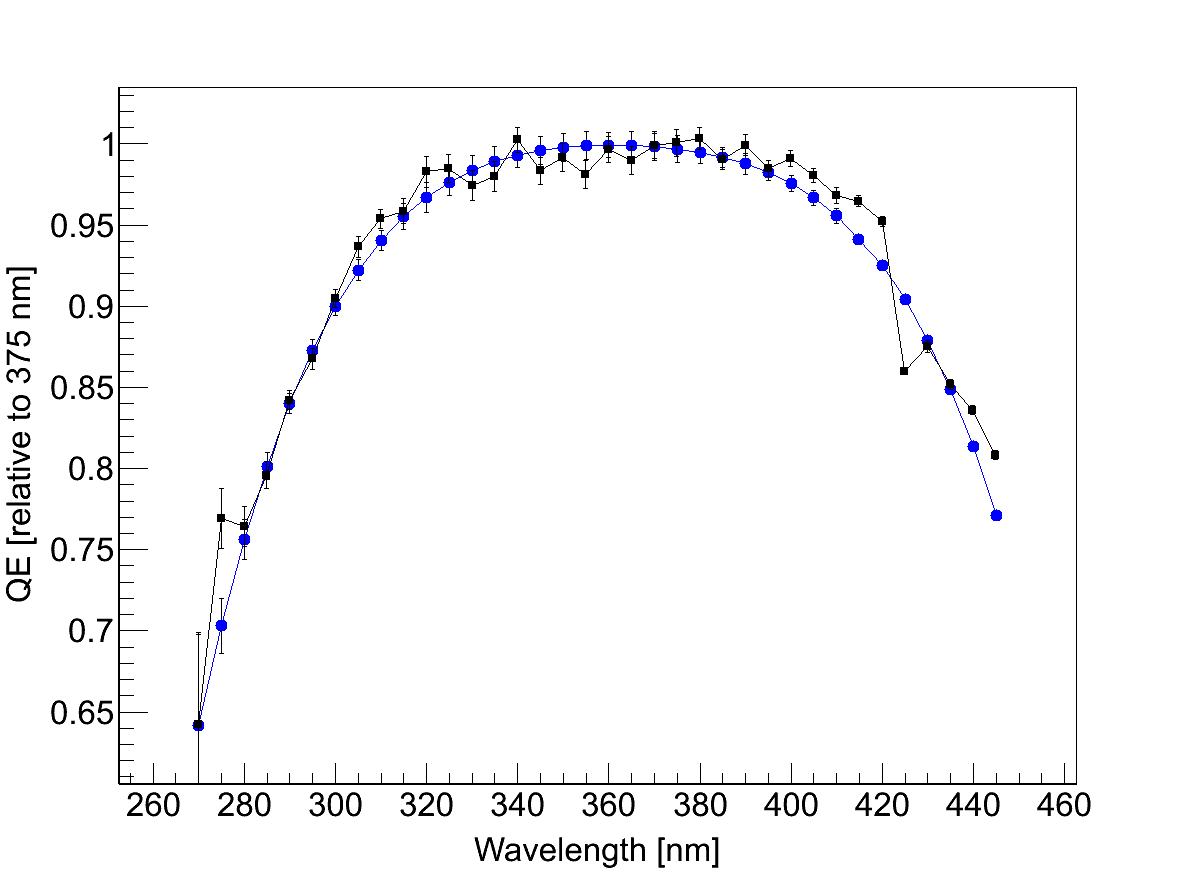}
  \caption{Shown in black squares is the measured relative Lab-PMT QE.
The error bars are the statistical uncertainty associated with the distribution of the response of the PMT at each wavelength. The blue circles are a fourth order polynomial fit to the data that serves to smooth out the measurement.}
  \label{fig:PMT_QE}
\end{figure}

Several measurements of the Lab-PMT QE were performed prior to and after the FD spectral calibration campaign, and these  measurements typically yielded curves consistent with the data shown as black squares in Figure~\ref{fig:PMT_QE}. The error bars are the statistical uncertainty associated with the spread in the response of the PMT to 100 pulses at each wavelength. 
The variations in the QE from point to point are typical when this kind of measurement is performed (e.g. see \cite{Rovero_MultiWave}), although they are not expected. In an attempt to smooth out these variations we fit the PMT QE curve with a fourth order polynomial shown as blue circles in the figure. The error bars in the fit are the relative statistical uncertainty for a given wavelength applied to the interpolated values in the fit. Deviations of the fit from measured points are largest at both the lower and upper ends of the wavelength range. However, the FD response is significant only in the range 310-410 nm (see Figure~\ref{fig:LMB4_Eff}) where the deviations are less than 2\% with RMS of approximately 1\%. We take this 1\% as a conservative estimate of the  systematic uncertainty in the measurement of the Lab-PMT QE: $\delta^\mathrm{Drum}_\mathrm{QESyst}(\lambda)\approx1\%$. 

Changing the nature of the fitted curve or using simply the measured black points from Figure~\ref{fig:PMT_QE} has little effect on measurements of EAS events.  For example a change of order 0.1\% on the reconstructed energy would result from using the measured QE points instead of the smoothed curve.  The small effect on energy occurs because in the region at high and low wavelengths where the fit deviates most from the measured points the FD efficiency is very low and the nitrogen fluorescence spectrum has no large features.

\subsection{Uncertainties in Lab Measurements}
\label{sec:Lab_Unc}
\begin{figure}[h]
 \centering
  \includegraphics[width=7.cm]{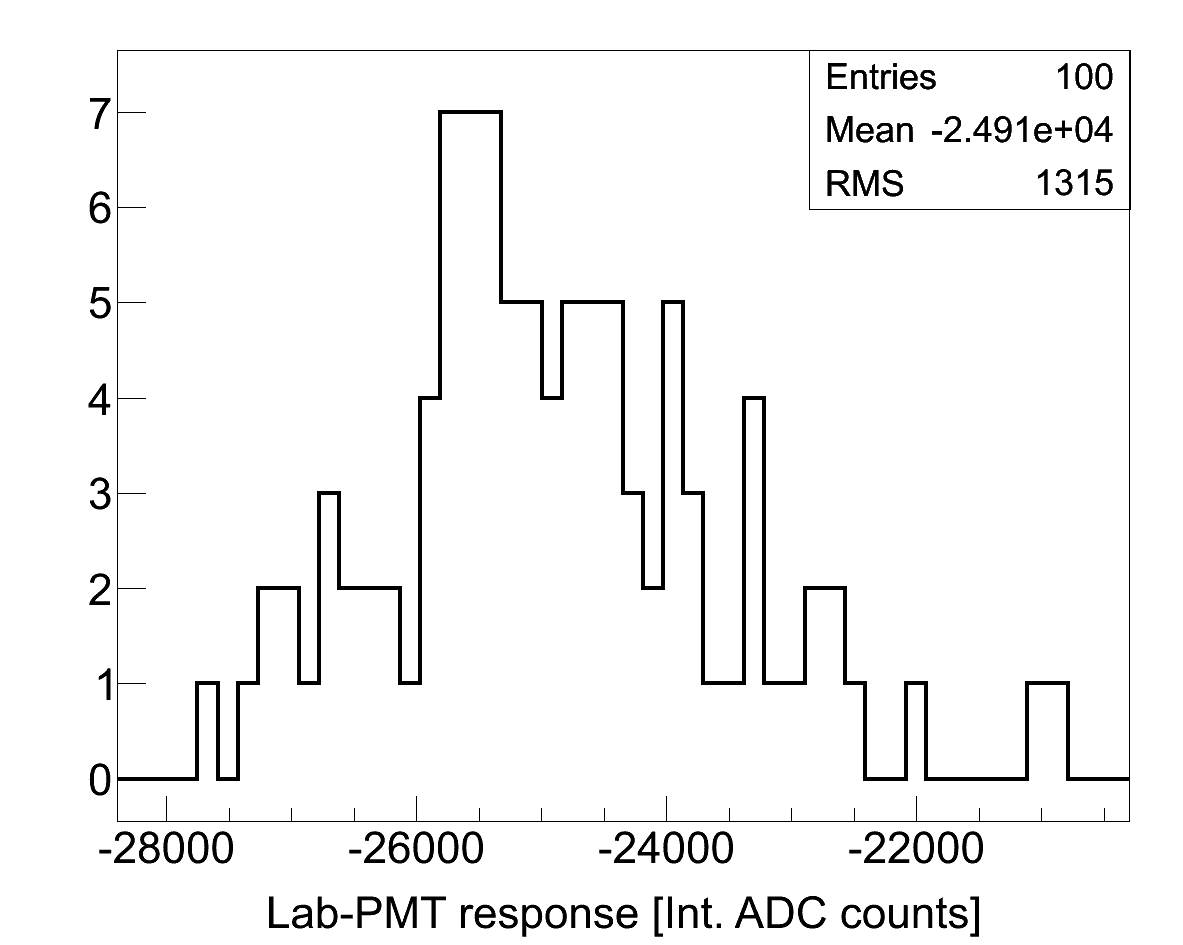}
  \caption{Distribution of the response of the Lab-PMT to 100 flashes of the drum at 375 nm.}
  \label{fig:LIT_Resp}
\end{figure}

The estimate of the statistical uncertainties for the various response distributions to the xenon flasher are taken as the standard deviation of the mean. Figure~\ref{fig:LIT_Resp} shows the response distribution of the Lab-PMT to 100 flashes of the drum at 375~nm where
$\delta^\mathrm{Drum}_\mathrm{PMTStat}(\lambda=375 \text{ nm})=\frac{\sigma(\lambda=375 \text{ nm})}{\sqrt{N}}\approx1\%$ of the average response $\overline{S^\mathrm{Drum}}(\lambda=375 \text{nm})$. The intensity at the monochromator output is known to be stable (with associated statistical uncertainties) through the monitoring PD spectra taken at the same time as the Lab-PMT data. 
A similar distribution was produced for each wavelength in the Lab-PMT QE measurement and gives $\delta^\mathrm{Drum}_\mathrm{QEStat}(\lambda)\approx1\%$.

\begin{figure}[h]
\centering
 \includegraphics[width=13.cm]{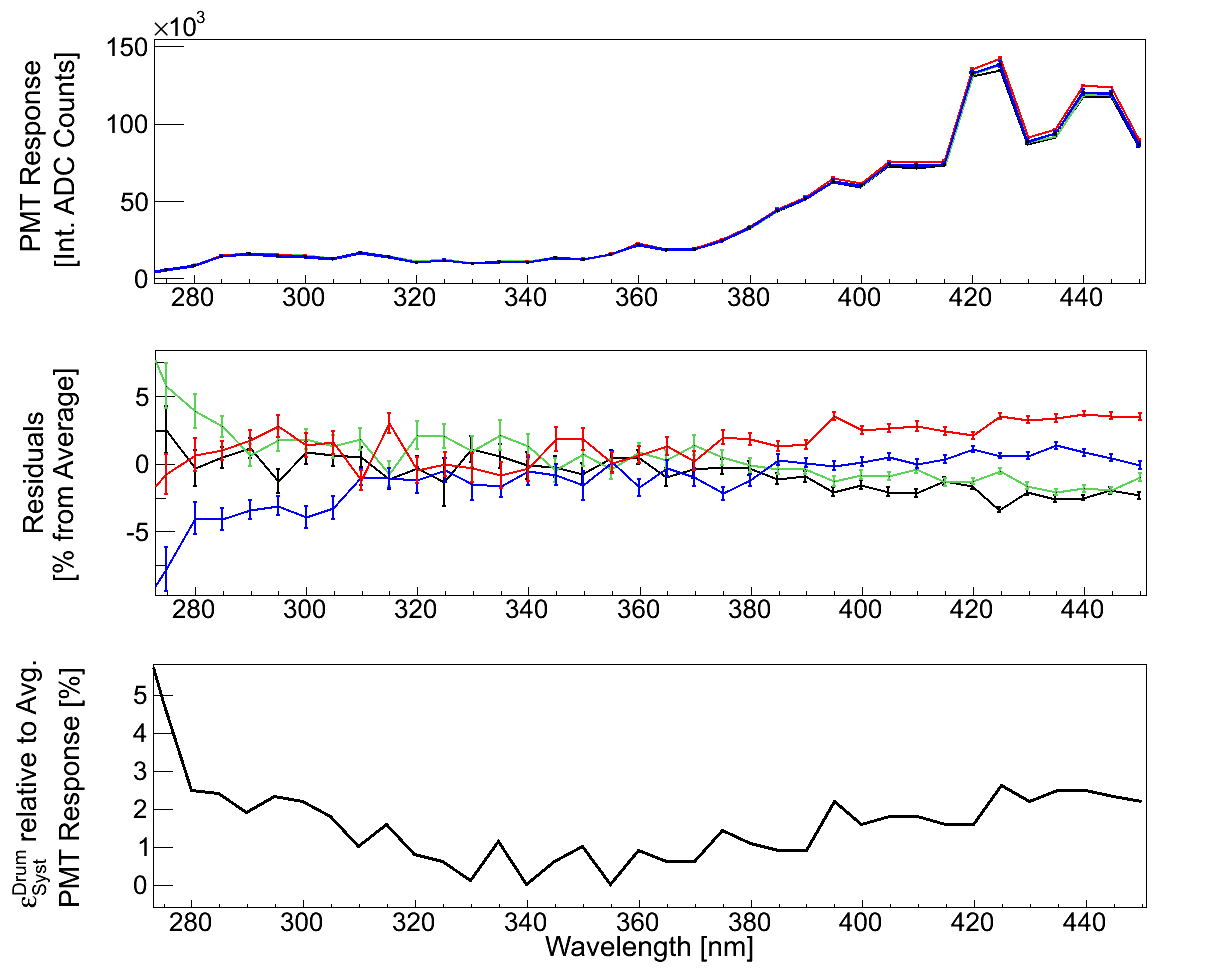}
  \caption{Four drum emission spectra as measured by the Lab-PMT (top), residuals from the average in percent (middle), the resulting systematic uncertainty $\varepsilon^\mathrm{Drum}_\mathrm{Syst}(\lambda)$ shown as a  percent of the PMT response at a given wavelength (bottom).}
  \label{fig:LIT_Syst}
\end{figure}

Estimating the systematic uncertainties associated with the relative drum emission spectrum is done by comparing the different drum emission spectra measured using the Lab-PMT over the course of the one-week campaign. Prior to the comparison, the Lab-PMT data are normalized by the simultaneous PD data at each wavelength to account for changes in the monochromator emission spectrum.  Shown in the top panel of Figure~\ref{fig:LIT_Syst} are the four drum spectra measured with the Lab-PMT that are used to calculate an average spectrum of the drum, and the middle plot shows the residuals from the average in percent as a function of wavelength.

Over most of the wavelength region where the FD efficiency is nonzero, 300~nm to 420~nm, the residuals plotted in Figure~\ref{fig:LIT_Syst} are close to agreement with each other within the statistical uncertainties. To estimate the systematic uncertainty of the drum emission at each wavelength we introduce an additive parameter, $\varepsilon^\mathrm{Drum}_\mathrm{Syst}(\lambda)$, such that calculating a $\chi^2$ per degree of freedom comparison via equation \eqref{eq:Lab_Syst} 
gives $\chi^2_\mathrm{ndf}\lesssim1$, and then this parameter is taken as the systematic uncertainty: 

\begin{equation}
 \chi^2_\mathrm{ndf}(\lambda)=\frac{1}{3}\sum_{n=1}^4\frac{\big(S^\mathrm{Drum}(\lambda)_n-\overline{S^\mathrm{Drum}}(\lambda)\big)^2}
{\big(\delta^\mathrm{Drum}_\mathrm{PMTStat}(\lambda)_n\big)^2+\big(\varepsilon^\mathrm{Drum}_\mathrm{Syst}(\lambda)\big)^2}\lesssim1 ~.
  \label{eq:Lab_Syst}
\end{equation}\

In equation~\eqref{eq:Lab_Syst} the Lab-PMT response (or drum emission) at a given wavelength is $S^\mathrm{Drum}(\lambda)$, the associated statistical uncertainty is $\delta^\mathrm{Drum}_\mathrm{PMTStat}(\lambda)$, and the average spectrum is $\overline{S^\mathrm{Drum}}(\lambda)$.

For a few wavelengths $ \chi^2_\mathrm{ndf}(\lambda) <1 $ without adding the systematic term in the denominator, and the corresponding systematic uncertainty is set to zero. But most wavelengths result in $ \chi^2_\mathrm{ndf}(\lambda)>1$ without the added term, so we calculate the systematic uncertainty for those wavelengths.   The result of this  procedure is that the non-zero Lab-PMT systematic uncertainties vary from  less than 1\% to approximately 3\%, and in the important region from 300 nm to 400 nm the average systematic uncertainty is, conservatively, about 1\%, see the bottom panel of Figure~\ref{fig:LIT_Syst}.

As a check, the PD spectra were treated with a similar evaluation of a systematic uncertainty at each wavelength as in equation~~\eqref{eq:Lab_Syst}. 
The corresponding systematic uncertainty  estimates for the PD would all be approximately 1\% or smaller. But there is no need to assess a systematic uncertainty on the drum intensity due to the PD since the PD data are only used to normalize the PMT data to reduce the spread in PMT measurements, and we use the spread in (normalized) PMT data for the systematic uncertainty. 

We estimate the overall systematic uncertainty on the intensity of the drum at each wavelength based on the QE measurement of the Lab-PMT ($\delta^\mathrm{Drum}_\mathrm{QESyst}(\lambda)~\approx1\%$ )  and the four measurements of the drum spectrum ($ \varepsilon^\mathrm{Drum}_\mathrm{Syst}(\lambda) ~\approx1\%$). Each of these uncertainties is conservatively about 1\% in the main region of the FD efficiency and nitrogen fluorescence spectrum, so a reasonable estimate of the overall systematic uncertainty of the drum intensity is found by adding them in quadrature: 1.4\%.

\section{FD Measurements}
\label{sec:FD_Meas}
During the March 2014 calibration campaign we measured the response of the eight telescopes, as specified in Table \ref{tab:FD_Comp}, in steps of 5~nm, over the course of five days. Data from the monitoring PD were also acquired during the FD measurements to be able to control for changes in the drum spectral emission. 
The procedure for measuring the telescope response to the multi-wavelength drum was to first scan from 255~nm to 445~nm in steps of 10~nm, and then scan from 250~nm to 450~nm in steps of 10~nm. At each wavelength a series of 100 pulses from the drum was recorded by the FD data acquisition at a rate of 1~Hz. A full telescope response is then an interleaving of these scans.  Later in analysis, wavelengths that result in essentially zero FD efficiency - at low and high wavelengths in the scan corresponding to the edges of the nitrogen spectrum -  were dropped and set to zero.   

In the previous sections we evaluated the systematic uncertainty in the drum light source intensity as a function of wavelength.  The contributions to this uncertainty are the spread in the four measurements of the drum intensity over the week of the calibration campaign and the systematic uncertainty in the quantum efficiency of the PMT used to measure the drum output.  

In this section we evaluate the uncertainty in the  responses of the telescopes  to the drum by comparing the responses of telescopes with the same optical components - see Table~\ref{tab:FD_Comp}. We do this comparison because we have not measured every telescope in the observatory, so we have to develop a single calibration constant for each wavelength for each of the four sets of optical components in the table. Then we use these calibration constants for all telescopes with like components (again, see Table~\ref{tab:FD_Comp}), including those not measured. Combining this uncertainty on the FD response, described below, with the drum emission systematic uncertainty will give the overall systematic uncertainties on the spectral calibration of the telescopes. As we will see below, of the four combinations of optical components in Table~\ref{tab:FD_Comp} three will result in systematics on FD response well below the systematics from the drum emission, but one pair of telescopes will have significantly different responses to the drum resulting in a systematic uncertainty larger than that from the drum intensity.

\subsection{FD Systematic Uncertainties Evaluated by Comparing Similar Telescopes}
\label{sec:FD_Sys_Unc_2}
We assume that the FDs built with like components - the same mirror and corrector ring types - should give similar responses, and to test that assumption we 
make a comparison between them to derive a meaningful systematic uncertainty. To that end we perform a $\chi^2$ test and introduce parameters to minimize the $\chi^2$ such that $\chi^2_\mathrm{ndf}\lesssim1$ for $ndf=34$, where there are 35 wavelengths used in the comparison. The parameters introduced are an overall scale factor $\beta$ that is applied to one of the FD responses, and then $\varepsilon_\mathrm{FD}$ which is an estimate of a systematic uncertainty that would be needed to account for the difference between the two telescopes.  Thus the raw response of one of the FDs as a function of wavelength is then $\beta*\big(FD_\mathrm{Resp}(\lambda)\pm\delta^{\mathrm{FD}}_\mathrm{Stat}(\lambda)\pm\varepsilon_\mathrm{FD}\big)$ in the comparison, where $\delta^\mathrm{FD}_\mathrm{Stat}(\lambda)$ is the statistical uncertainty (small) as mentioned at the end of this section. The scale factor $\beta$ does not represent a systematic uncertainty, it just  accounts for any overall difference in response between the two telescopes. This is similar to performing a relative calibration analysis as in~\cite{Rel_Cal_Monitor_FD} between the two telescopes.   

The minimization is done in two steps according to equation~\eqref{eq:FD_Chi_Test} where the sum is over the $N_{\lambda}$ measured wavelength points:

\begin{equation}\label{eq:FD_Chi_Test}
\chi^2_\mathrm{ndf}=\frac{1}{34}\sum_\mathrm{n=1}^\mathrm{N_{\lambda}}\frac{\Big(FD_1(\lambda)_\mathrm{n}-\beta*FD_2(\lambda)_\mathrm{n}\Big)^2}{\Big(\delta^\mathrm{FD_1}_\mathrm{Stat}
(\lambda)_\mathrm{n}\Big)^2+\Big(\beta*\delta^\mathrm{FD_2}_\mathrm{Stat}(\lambda)_\mathrm{n}\Big)^2+\Big(\beta*\varepsilon_\mathrm{FD}\Big)^2}{} ~.
\end{equation}
First a minimum in $\chi^2$ is found by setting $\varepsilon_\mathrm{FD}=0$ and allowing the scale factor~$\beta$ to vary.
Once $\beta$ has been determined, $\varepsilon_\mathrm{FD}$ is allowed to vary until $\chi^2_\mathrm{ndf}\lesssim1$.
Prior to the minimization the $FD_2(\lambda)$ response data are normalized by the ratio of the monitoring PD response as measured at $FD_1$ and $FD_2$ for a given wavelength. This serves to divide out any change in intensity of the light source as measured by the PD just downstream of the monochromator, and this normalization does, as expected, improve the agreement in response for some telescope pairs. 

\begin{table}[h]
\caption{$\varepsilon_\mathrm{FD}$ and $\beta$ values obtained via equation~\eqref{eq:FD_Chi_Test} for the similarly constructed telescopes. The $\varepsilon_\mathrm{FD}$ for a given pair of telescopes is given in percentage relative to the averaged response of the pair of telescopes at 375~nm.} 
\begin{center}
\begin{tabular}{|l|r|r|}
\hline
\textbf{FDs} & $\varepsilon_\mathrm{FD}$ [\%] & $\beta$\\
\hline
\hline
Coihueco 2/3           &  0.34  &  0.97\\\hline
Coihueco 4/5           &  0.48  &  1.02\\\hline
Los Morados 4/5        &  0.14  &  1.01\\\hline
Los Leones 3/4         &   1.7  &  1.05\\\hline
\end{tabular}
\end{center}
\label{tab:FD_Syst_Vals}
\end{table}

The values for $\varepsilon_\mathrm{FD}$ and $\beta$ are listed in Table~\ref{tab:FD_Syst_Vals} for each pair of telescopes that are constructed with nominally identical components, and the systematic uncertainties, $\varepsilon_\mathrm{FD}$, are reported as a percentage of the average response of the two telescopes at 375~nm. 

Aside from Los Leones telescopes 3 and 4, the $\varepsilon_\mathrm{FD}$ values derived through this minimization technique are all less than 0.5~\% , and the $\beta$ scale factors are all within about 3~\% of unity.
The values obtained for Los Leones, although larger than the others, are still small. By trying to find a reason for this difference we note that telescope 4 was part of the Engineering Array (EA, \cite{EA2004}) together with telescope 5. However, both telescopes were rebuilt after the EA operation, particularly the mirrors were all replaced by new ones after re-setting the design parameters. So, the discrepancy between LL 3 and 4 is highly probably not caused by any difference in used materials and, in any case, is included in the uncertainties.

We use the $\varepsilon_\mathrm{FD}$ calculated for a given pair of FD telescopes as a systematic uncertainty across all wavelengths for all telescopes of the corresponding construction; see Table~\ref{tab:FD_Comp}. These systematic uncertainties are then normalized by the telescope response at 375 nm and are added in quadrature with the uncertainties associated with the spread in Lab-PMT measurements of the drum (about 1\% in important wavelength range, a function of wavelength), and the Lab-PMT QE (1\%, not wavelength dependent) to calculate the overall systematic uncertainty on telescopes of each combination of optical components.  An example result is plotted as the red brackets in Figure~\ref{fig:LMB4_Eff} for the Los Morados telescopes 4 and 5; for this pair (and like telescopes) the overall systematic uncertainty on the FD response is approximately $\sqrt{1^2+1^2+0.14^2}=1.4 \%$, and it is dominated by the uncertainty in the drum intensity. For the Coihueco instruments the overall systematic uncertainty is about 1.5\%.   For the telescopes  at Los Leones the uncertainty from the  difference in response between the two telescopes is larger than the drum-related systematic uncertainties, and the overall systematic uncertainty on all of the Los Leones telescopes is about 2.2\%.

The statistics of the data taken with the drum light source at the FD telescopes also contribute to the uncertainties on the calibration constants. The typical spread in the average response of the 440 PMTs to the 100 drum pulses at a given wavelength is 0.4\% RMS, which is much smaller than the systematic uncertainties. Adding the statistical uncertainty in quadrature with the systematic uncertainties yields the overall uncertainties on the calibration constants listed in Table~\ref{tab:FD_Overall_Uncertainties}, which are the main result of this work.

\begin{table}[h]
\caption{Overall uncertainties on spectral calibration constants for the pairs of telescopes measured and all other (unmeasured) telescopes with the same optical components.} 
\begin{center}
\begin{tabular}{|l|c|r|
}
\hline
\textbf{FDs} &  \textbf{Overall} & \textbf{FDs with same}\\
~ & \textbf{uncertainty [\%]} & \textbf{components} \\
\hline
\hline
Coihueco 2/3           &  1.5  &  CO2/3\\\hline
Coihueco 4/5           &  1.5  &  CO1,4-6, LA, HEAT\\\hline
Los Morados 4/5        &  1.5  &  LM\\\hline
Los Leones 3/4         &   2.2
  &  LL1-6\\\hline
\end{tabular}
\end{center}
\label{tab:FD_Overall_Uncertainties}
\end{table}

\subsection{Photodiode Monitor Data}
We performed a comparison between the average dark hall PD spectrum and each of the spectra measured for the data-taking nights at the FDs to ensure that the light source was stable and was consistent with what had been measured in the lab. An overall correction of $1.00\pm0.01$ night to night was found as the average ratio of the PD response at the FD to that at the lab to accommodate any overall variations in intensity or response due to temperature effects, and then we performed a $\chi^2$ comparison for all the measured wavelengths. For all measuring nights at the FDs the PD spectra agree very well, the comparison gives a $\chi^{2}_\mathrm{ndf}\sim1$ where $ndf=34$ for each, implying that the spectrum as observed by the PD was the same at all locations. 

\section{Calculation of the FD Efficiency}
\label{sec:FD_Eff}
We calculate the relative FD efficiency for each telescope by dividing the measured telescope response to the drum by the measured drum emission spectrum. The relative drum emission spectrum is measured as described in section~\ref{sec:Lab_Meas} and takes into account the Lab-PMT quantum efficiency over the  range from 250 nm to 450 nm.

\begin{figure}[h!]
\centering
 \includegraphics[width=12.cm]{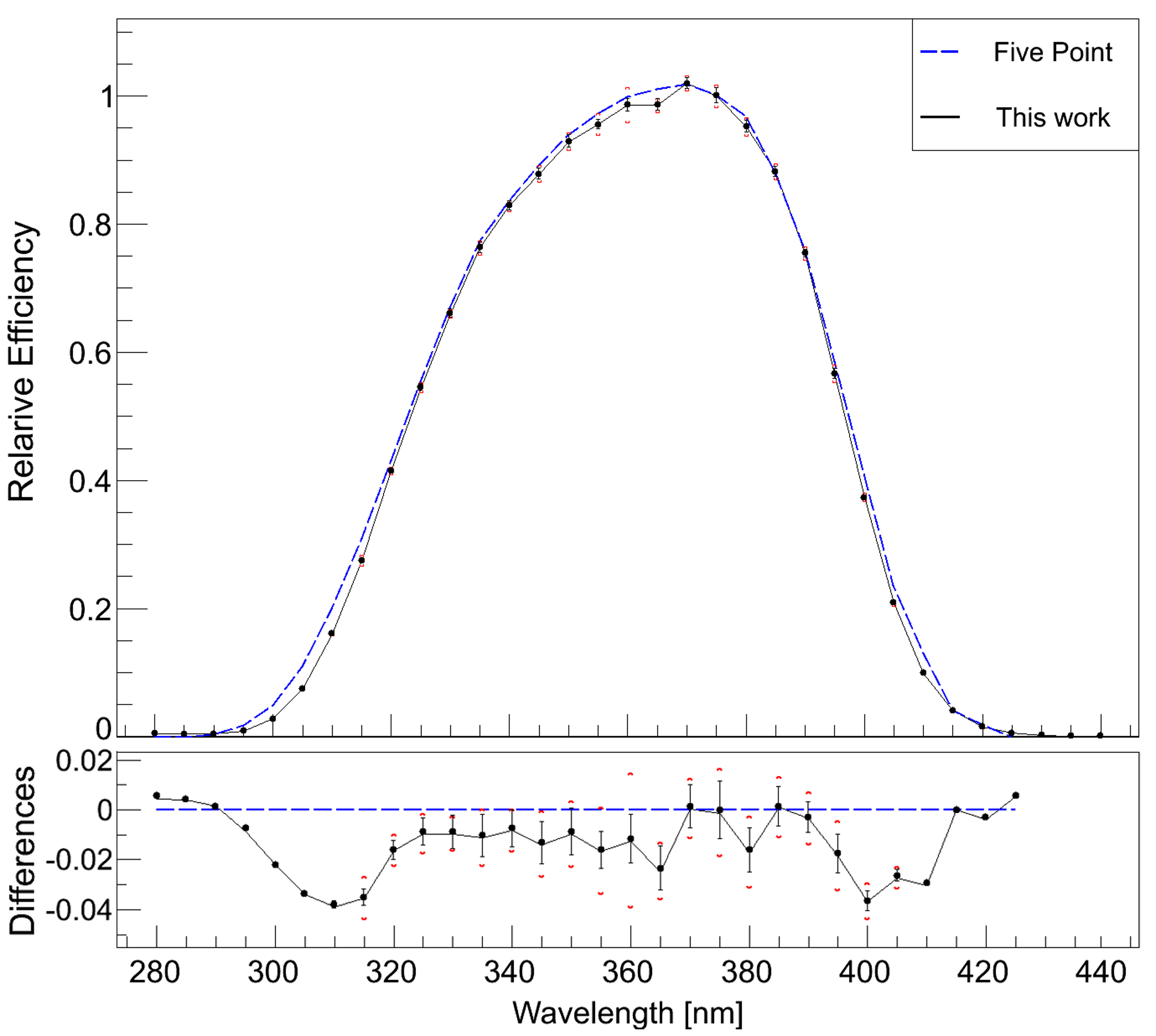}
  \caption{Relative efficiencies for the average of Los Morados telescopes 4 and 5. {\it Top}:  the five filter curve shown as a dashed  blue line~\cite{Rovero_MultiWave} and the monochromator result shown as the solid line. Error bars are statistical uncertainties, and the red brackets are the systematic uncertainties calculated as described in section~\ref{sec:FD_Meas}. {\it Bottom}: difference between the five filter result and this work. The error bars and brackets are the same as in the top plot, shown here for clarity.}
  \label{fig:LMB4_Eff}
\end{figure}

The relative efficiency for a given telescope at a given wavelength, $FD_\mathrm{eff}^\mathrm{Rel}(\lambda)$, is calculated for each wavelength from 280~nm to 440~nm in steps of 5~nm:
\begin{equation}
 FD_\mathrm{eff}^\mathrm{Rel}(\lambda)=\frac{FD_\mathrm{Resp}(\lambda)*{QE^\mathrm{Lab}_\mathrm{PMT}}(\lambda)}{\overline{S^\mathrm{Drum}}(\lambda)}*\frac{1}{FD_\mathrm{eff}(\lambda=375~\text{nm})}
\label{eq:Eff_Calc}{} ~.
\end{equation}
The curves are taken relative to the efficiency of the telescope at 375~nm since this is what is used in the Pierre Auger Observatory reconstruction software~\cite{argiro2007offline} for all FD calculations. The  range in wavelength from 280~nm to 440~nm  used for evaluating the FD efficiency is smaller than the range measured in the lab because below 280~nm and above 440 nm the light level is near zero intensity for the nitrogen emission spectrum and the FD response is also very near zero. As an example, 
Figure~\ref{fig:LMB4_Eff} shows the relative efficiency for the Los Morados telescopes 4 and 5 based on this work compared with the previous measurement~\cite{Rovero_MultiWave}.

The uncertainties in the FD efficiencies have statistical and systematic components associated with the measurement of the relative emission spectrum of the drum, the Lab-PMT QE, and the FD response to the multi-wavelength drum. 
The statistical uncertainties associated with the lab work and the FD responses are propagated through the calculation of the FD efficiency via equation~\eqref{eq:Eff_Calc} as a function of wavelength.
All systematic uncertainties described above associated with the lab work, the Lab-PMT and its QE, the FD response, and $\varepsilon_\mathrm{FD}$ for a given FD telescope, are added together in quadrature as a function of wavelength. 

For much of the wavelength range the new results agree with the older five-point scan. The disagreement at the shortest and longest wavelengths is perhaps not surprising since the previous lowest and highest measurements were at 320~nm and 405~nm, and the efficiency was extrapolated to zero from those points following the piecewise curve~\cite{matthews2003optical}. The efficiency was assumed to go to zero below 295~nm and above 425~nm.

\section{Comparison of telescopes with differing optical components}
\label{sec:discus}
After estimating the systematic uncertainties for each measured FD telescope, $\varepsilon_\mathrm{FD}$, we made a $\chi^2$ comparison between the six combinations of unlike FD optical components listed in Table~\ref{tab:FD_Comp} to determine whether the unlike components result in any different telescope responses. In calculating the 
$\chi^2_\mathrm{ndf}$ for the differently constructed FD telescopes we use the ratio of the PD data taken at the corresponding FDs to normalize the average response of one of the FD types. The PD data from the two FD data-taking nights are averaged  as a function of wavelength and the ratio of the PD averages from the two types of FDs are applied to the combined FD response along with the statistical and systematic uncertainties. Using this normalization serves to divide out any differences in the drum spectrum between the two measurements of the FD types. An example for calculating the $\chi^2$ between the average of Coihueco  2 and Coihueco 3 ($\overline{S_\mathrm{CO23}(\lambda)_\mathrm{n}}$) and the average of Coihueco  4 and  5 follows. The uncertainties in equation \ref{eq:Diff_FD_Comp} have obvious labels; for example $\varepsilon_\mathrm{FD}^\mathrm{CO23}$ is the systematic uncertainty for the Coihueco telescopes 2 and 3 from Table~\ref{tab:FD_Syst_Vals}.

\begin{equation}
\begin{split}
&\chi^{2}_\mathrm{ndf}=\frac{1}{34}\sum_\mathrm{n=1}^\mathrm{N_\lambda}\frac{\Big(\overline{S_\mathrm{CO23}(\lambda)_\mathrm{n}}-PD_\mathrm{Ratio}(\lambda)_\mathrm{n}*\overline{S_\mathrm{CO45}(\lambda)_\mathrm{n}}\Big)^2}{\big(\delta^\mathrm{CO23}_\mathrm{Stat}(\lambda)_\mathrm{n}\big)^2+\big(PD_\mathrm{Ratio}(\lambda)_\mathrm{n}*\delta^\mathrm{CO45}_\mathrm{Stat}
(\lambda)_\mathrm{n}\big)^2 + \big(\varepsilon_\mathrm{FD}^\mathrm{CO23}\big)^2  + \big(\varepsilon_\mathrm{FD}^\mathrm{CO45}\big)^2} ~, \\
&  \\
&PD_\mathrm{Ratio}(\lambda)\equiv\frac{\overline{S^\mathrm{PD}_\mathrm{CO23}}(\lambda)}{\overline{S^\mathrm{PD}_\mathrm{CO45}}(\lambda)} ~.
\end{split}
\label{eq:Diff_FD_Comp}
\end{equation}

The results of the comparisons are shown in Table~\ref{tab:Chi_Comp}.  The telescopes with different components are all significantly different from each other except when comparing the average of Los Leones  telescopes 3  and 4 to the average of Los Morados telescopes 4 and 5. In principle this low   $\chi^2$ could indicate that all telescopes constructed with components like those at Los Leones 3 and 4 and those constructed like Los Morados telescopes 4 and 5 have the same response, and therefore could share the relative calibration constants that are the goal of this work. However, the two detectors at Los Leones have a much greater difference in response between them than do the two telescopes from Los Morados: the systematic uncertainty in Table~\ref{tab:FD_Syst_Vals} for Los Leones telescopes is more than a factor of 10 larger than that for Los Morados ones. The large systematic uncertainty for the telescopes at Los Leones could be masking a real difference with those at Los Morados. For this reason we feel it is reasonable not to combine the Los Leones and Los Morados telescopes to calculate the final spectral calibration constants.

We conclude that all four sets of FD telescopes listed in Table~\ref{tab:FD_Comp} need different spectral calibrations, and  four sets of calibration constants have been computed.

Examining the results in Table~\ref{tab:Chi_Comp} and Table~\ref{tab:FD_Comp} we note that the largest $\chi^2_\mathrm{ndf}$ values in Table~\ref{tab:Chi_Comp} are associated with changing mirrors not changing corrector rings. For example, comparing Coihueco 4/5 with Los Morados 4/5 changes only the mirror and gives a $\chi^2_\mathrm{ndf}$ of 55, but comparing Coihueco 4/5 with Coihueco 2/3, which changes only the corrector ring, yields a $\chi^2_\mathrm{ndf}$ of 5.6.  Changing both components by comparing Coihueco 2/3 with Los Morados 4/5 gives a $\chi^2_\mathrm{ndf}$ of 161, but we note that the telescopes at Los Morados have a very small systematic uncertainty in Table~\ref{tab:FD_Syst_Vals}. These examples have so far left out the Los Leones telescopes. The large systematic uncertainty derived by comparing the two Los Leones telescopes reduces the $\chi^2_\mathrm{ndf}$ values when comparing to other telescopes, but the idea that the mirrors are the main effect is still present when comparing the Los Leones telescopes to the others.

\begin{table}[h]
\caption{Comparison of spectral response for FD telescopes with different components. $\chi^2_\mathrm{ndf}$ values obtained for the sets in Table \ref{tab:FD_Comp}, where $ndf=34$.} 
\begin{center}
\begin{tabular}{|l|r|}
\hline
\textbf{Comparison} &$\mathbf{\chi^2_\mathrm{ndf}}$\\
\hline
\hline
Coihueco 2/3 vs. Coihueco 4/5           &  2.4  \\\hline
Los Morados 4/5 vs. Los Leones 3/4      &  0.21  \\\hline
Coihueco 4/5 vs Los Morados 4/5         &  57  \\\hline
Coihueco 2/3 vs. Los Leones 3/4         &  10  \\\hline
Coihueco 2/3 vs. Los Morados 4/5        &  144 \\\hline
Los Leones 3/4 vs. Coihueco 4/5         &  6.7 \\\hline
\end{tabular}
\end{center}
\label{tab:Chi_Comp}
\end{table}

\section{Effect on Physics Measurables}
\label{sec:Physics}

To evaluate the effect a new calibration has on physics measurables, we reconstructed a set of events using the new calibration and compare to results from that same set of events using the prior calibration. When we did this exercise upon changing from initial piecewise to the five-point calibration, the reconstructed energies increased about 4\% at $10^{18}$ eV, and the increase lessened slightly to 3.6\% at $10^{19}$ eV  \cite{All_Auger_ICRC_2013}. The lessening of the energy increase due the five-point calibration is understood because much of the change in calibration was at low wavelengths, and the five-point calibration makes the FDs less efficient at short wavelengths making the reconstructed energy higher. The higher energy events make more light, and they can be detected at greater distances than lower energy events. But at greater distances more of the short wavelength light will be Rayleigh scattered away, so the lower wavelengths - and the change in calibration there - do not affect the higher energy events as much when we change to the new calibration.  

When we change from the five-point to the calibration described here, the reconstructed energies increase on average over all FD telescopes by about 1\%, and that increase is relatively flat in energy. However, this increase is not the same at all the telescopes. The increase in reconstructed energy is greatest at Los Leones, about 2.8\% at $10^{18}$ eV falling to 2.5\% at $10^{19}$ eV. For Los Morados the reconstructed event energies increase by about 1.8\% without much energy dependence. For all other telescopes the energy increases, but those increases are less than 0.35\% for all energies.

All these changes in the reconstructed energy are important to know to fully characterize the telescopes. Regarding the associated uncertainties, they are all significantly smaller than the uncertainties involved in the energy scale for the FD telescopes (see Table 3 in~\cite{pierre2015pierre}), particularly the 3.6\% from the Fluorescence yield and the 9.9\% from the FD calibration.

\section{Conclusions}
\label{sec:Conc}
Determining the spectral response of the Pierre Auger Observatory fluorescence telescopes is essential to the success of the experiment. A method using a monochromator-based portable light source has been used  for eight FD telescopes with measurements performed every five nanometers from 280~nm to 440~nm. With the calibration of these eight telescopes, the four possible combinations of different optical components in the FD were covered, thus assuring the spectral calibration of all FD telescopes at the observatory. 

The uncertainty associated with the emission spectrum of the drum light source used for the calibration was
found to be 1.4\%, which is an improvement on our previous 3.5\%~\cite{Rovero_MultiWave}. 

For the present work we compared telescopes with nominally the same optical components, and we find that such pairs have the same spectral response within a fraction of a percent - as expected - for three out of the four pairs of like telescopes.  But one pair with like components, the oldest telescopes in the observatory, shows a significant difference in spectral response.   

The overall uncertainty in the FD spectral response is 1.5\% for 21 of the 27 telescopes. The overall systematic uncertainty for the remaining six telescopes is 2.2\%, and is  somewhat larger on account of  the larger difference between the two telescopes measured. 

We also compared the differently constructed telescopes. These comparisons show significantly different efficiencies as a function of wavelength, with differences mainly in the rising edge of the efficiency curve between 300~nm and 340~nm. The differences seem to come mostly from the two different mirror types, and they are reflected in different calibration constants for each of the four combinations of optical components.  

The new calibration constants affect the reconstruction of EAS events, and we looked at two important quantities.  The primary cosmic ray energy increases by 1.8\% to 2.8\% for half of the telescopes in the observatory, and for the  other half the change in energy is negligible. The position of the maximum in shower development in the atmosphere, $X_\mathrm{max}$, is not changed significantly by the change in calibration.

\section*{Acknowledgments}

\begin{sloppypar}
The successful installation, commissioning, and operation of the Pierre Auger Observatory would not have been possible without the strong commitment and effort from the technical and administrative staff in Malarg\"ue. We are very grateful to the following agencies and organizations for financial support:
\end{sloppypar}

\begin{sloppypar}
Argentina -- Comisi\'on Nacional de Energ\'\i{}a At\'omica; Agencia Nacional de Promoci\'on Cient\'\i{}fica y Tecnol\'ogica (ANPCyT); Consejo Nacional de Investigaciones Cient\'\i{}ficas y T\'ecnicas (CONICET); Gobierno de la Provincia de Mendoza; Municipalidad de Malarg\"ue; NDM Holdings and Valle Las Le\~nas; in gratitude for their continuing cooperation over land access; Australia -- the Australian Research Council; Brazil -- Conselho Nacional de Desenvolvimento Cient\'\i{}fico e Tecnol\'ogico (CNPq); Financiadora de Estudos e Projetos (FINEP); Funda\c{c}\~ao de Amparo \`a Pesquisa do Estado de Rio de Janeiro (FAPERJ); S\~ao Paulo Research Foundation (FAPESP) Grants No.\ 2010/07359-6 and No.\ 1999/05404-3; Minist\'erio de Ci\^encia e Tecnologia (MCT); Czech Republic -- Grant No.\ MSMT CR LG15014, LO1305, LM2015038 and CZ.02.1.01/0.0/0.0/16\_013/0001402; France -- Centre de Calcul IN2P3/CNRS; Centre National de la Recherche Scientifique (CNRS); Conseil R\'egional Ile-de-France; D\'epartement Physique Nucl\'eaire et Corpusculaire (PNC-IN2P3/CNRS); D\'epartement Sciences de l'Univers (SDU-INSU/CNRS); Institut Lagrange de Paris (ILP) Grant No.\ LABEX ANR-10-LABX-63 within the Investissements d'Avenir Programme Grant No.\ ANR-11-IDEX-0004-02; Germany -- Bundesministerium f\"ur Bildung und Forschung (BMBF); Deutsche Forschungsgemeinschaft (DFG); Finanzministerium Baden-W\"urttemberg; Helmholtz Alliance for Astroparticle Physics (HAP); Helmholtz-Gemeinschaft Deutscher Forschungszentren (HGF); Ministerium f\"ur Innovation, Wissenschaft und Forschung des Landes Nordrhein-Westfalen; Ministerium f\"ur Wissenschaft, Forschung und Kunst des Landes Baden-W\"urttemberg; Italy -- Istituto Nazionale di Fisica Nucleare (INFN); Istituto Nazionale di Astrofisica (INAF); Ministero dell'Istruzione, dell'Universit\'a e della Ricerca (MIUR); CETEMPS Center of Excellence; Ministero degli Affari Esteri (MAE); Mexico -- Consejo Nacional de Ciencia y Tecnolog\'\i{}a (CONACYT) No.\ 167733; Universidad Nacional Aut\'onoma de M\'exico (UNAM); PAPIIT DGAPA-UNAM; The Netherlands -- Ministerie van Onderwijs, Cultuur en Wetenschap; Nederlandse Organisatie voor Wetenschappelijk Onderzoek (NWO); Stichting voor Fundamenteel Onderzoek der Materie (FOM); Poland -- National Centre for Research and Development, Grants No.\ ERA-NET-ASPERA/01/11 and No.\ ERA-NET-ASPERA/02/11; National Science Centre, Grants No.\ 2013/08/M/ST9/00322, No.\ 2013/08/M/ST9/00728 and No.\ HARMONIA 5--2013/10/M/ST9/00062, UMO-2016/22/M/ST9/00198; Portugal -- Portuguese national funds and FEDER funds within Programa Operacional Factores de Competitividade through Funda\c{c}\~ao para a Ci\^encia e a Tecnologia (COMPETE); Romania -- Romanian Authority for Scientific Research ANCS; CNDI-UEFISCDI partnership projects Grants No.\ 20/2012 and No.194/2012 and PN 16 42 01 02; Slovenia -- Slovenian Research Agency; Spain -- Comunidad de Madrid; Fondo Europeo de Desarrollo Regional (FEDER) funds; Ministerio de Econom\'\i{}a y Competitividad; Xunta de Galicia; European Community 7th Framework Program Grant No.\ FP7-PEOPLE-2012-IEF-328826; USA -- Department of Energy, Contracts No.\ DE-AC02-07CH11359, No.\ DE-FR02-04ER41300, No.\ DE-FG02-99ER41107 and No.\ DE-SC0011689; National Science Foundation, Grant No.\ 0450696; The Grainger Foundation; Marie Curie-IRSES/EPLANET; European Particle Physics Latin American Network; European Union 7th Framework Program, Grant No.\ PIRSES-2009-GA-246806; European Union's Horizon 2020 research and innovation programme (Grant No.\ 646623); and UNESCO.
\end{sloppypar}

\vspace{2cm}

\bibliography{references}

\end{document}